\documentclass[english,aps,manuscript]{revtex4}
\usepackage[T1]{fontenc}
\usepackage[latin9]{inputenc}
\usepackage{amssymb}

\makeatletter
\@ifundefined{textcolor}{}
{%
 \definecolor{BLACK}{gray}{0}
 \definecolor{WHITE}{gray}{1}
 \definecolor{RED}{rgb}{1,0,0}
 \definecolor{GREEN}{rgb}{0,1,0}
 \definecolor{BLUE}{rgb}{0,0,1}
 \definecolor{CYAN}{cmyk}{1,0,0,0}
 \definecolor{MAGENTA}{cmyk}{0,1,0,0}
 \definecolor{YELLOW}{cmyk}{0,0,1,0}
}

\makeatother

\usepackage{babel}
\begin{document}

\title{Comment to the paper: 'Seeded quantum FEL at 478 keV' AIP Conf. Proc.
'Light at Extreme Intensities' 1462 173-176 (2012)'}

\author{V. Petrillo}

\email{petrillo@mi.infn.it}

\selectlanguage{english}%

\affiliation{$ $INFN-Università degli Studi Milano ,Via Celoria, 16 20133 Milano,
Italy}
\begin{abstract}
We criticize the thesis exposed the paper 'Seeded quantum FEL at 478
keV' AIP Conf. Proc. 'Light at Extreme Intensities,' 1462 173-176
(2012), which presents the possibility of producing gamma rays at
478 KeV by means of a seeded quantum FEL driven by an electron beam
at 125 MeV , current I=40 A, interacting with an infrared laser. We
show that, in the case analyzed, the FEL Pierce parameter has a value
two orders of magnitude less than what claimed in the paper in question,
overturning the conclusions of the analysis.
\end{abstract}
\maketitle
In the paper \cite{geuente}, the Authors claim that a seeded quantum
optical FEL driven by an electron beam at 125 MeV of energy, with
current I=40 A, transverse dimension $\sigma_{x}=5-50\mu m$ and Twiss
parameter $\beta*=10$cm, interacting with a laser beam at wavelength
$\lambda_{L}=1\mu m$ and with a rather low laser parameter $a_{0}=0.03$
can produce gamma rays at 478 keV with an expected spectral flux up
to $10^{11}eV^{-1}s^{-1}$. The statements of this paper violate not
only the physical laws, but also the common sense. The authors in
fact base their estimates on an manipulated expression of the classical
Pierce parameter $\rho$ (that then appears in the expression for
the quantum parameter $\bar{\rho},$ ruling the efficiency of their
quantum FEL) given by\cite{robbPP}: 
\begin{equation}
\rho_{Diffr}=\left[\frac{I}{\gamma I_{A}}(\frac{f_{c}^{2}K^{2}}{1+K})\right]{}^{1/2}
\end{equation}

obtaining the value $\rho_{Diffr}=9\,10^{-5}$.

Unfortunately, they completely ignore that the previous formula is
valid in cases of diffraction dominated FELs and derives from:

\begin{equation}
\rho_{Diffr}=\rho(\frac{\sigma_{x}}{\sigma_{rad}})^{2/3}
\end{equation}

where $\rho$ is the usual 1D Pierce parameter (as defined for instance,
in the paper by Ming-Xie \cite{key-4}) and $\sigma_{rad}$ the transverse
dimension of the radiation.

As can be easily seen, eq. (2), when correctly used for diffraction
dominated FELs, $\sigma_{rad}>\sigma_{x}$ and consequently $\rho_{Diffr}<\rho$,
translates the obvious consideration that diffraction can never strengthen
the radiation, but only oppose to it.

In the case treated by \cite{geuente}, the radiation diffraction
is completely negligible, due to the value of the radiation wavelength
in the gamma rays range, and surely the Rayleigh length of the radiation
is very much larger than the Twiss electron beta function, so that
the usual expression of $\rho$ must be used.

The usual 1D Pierce parameter $\rho$ (see for instance, the paper
by Ming-Xie \cite{key-4}) with the same parameters of \cite{geuente},
in the most advantageous of their cases, gives:

\[
\rho=1.5\,10^{-6}
\]

almost two orders of magnitude less than what they claim, changing
completely the conclusions of their study.

An energy spread $\Delta\gamma/\gamma$ of the order of $10^{-3}$
or even $10^{-4}$, in fact, does not satisfy one of the basic criteria
for the lasing, i.e. $\Delta\gamma/\gamma\lesssim\rho$. 

The production of brilliant X rays with optical FELs requires electron
beams with current and brightness very much larger than those proposed
in \cite{geuente} as shown for instance in \cite{key-5}or \cite{key-6},
and summarized in \cite{key-7}, with suitable emittances and energy
spreads and laser energy. 

In conclusion, the use of an expression of the Pierce parameter applied
outside its validity range leads M. M. Guenther, D. Habs et al.  to
claim that it is possible to obtain a brilliant pulse of gamma rays
with an optical FEL, contrarily instead to what found within a correct
formulation of the problem.

\end{document}